\documentclass[twocolumn]{aastex61}

\shorttitle{A study on the V Gru binary system}
\usepackage{lineno}
\usepackage{amsmath}
\usepackage{graphicx}
\usepackage{xcolor}
\usepackage{longtable}
\usepackage{times}
\usepackage{tablefootnote}
\usepackage{appendix}
\usepackage{lineno}

\usepackage{soul}

\begin{document}

\title{The First Multiband Photometric Light Curve Solutions of the V Gru Binary System from the Southern Hemisphere}

\correspondingauthor{Atila Poro}
\email{$poroatila@gmail.com$}

\author{Mehmet TANRIVER}
\affil{Astronomy and Space Science, Science Faculty, University of Erciyes, Kayseri 38039, Türkiye}
\affil{Erciyes University, Astronomy and Space Science Observatory Application and Research Center, Kayseri 38039, Türkiye}

\author{Atila Poro}
\affil{Astronomy Department of the Raderon AI Lab., BC., Burnaby, Canada}

\author{Ahmet BULUT}
\affil{Department of Physics, Faculty of Arts and Sciences, Çanakkale Onsekiz Mart University, Terzioğlu Kampüsü, TR-17020, Çanakkale, Türkiye}
\affil{Astrophysics Research Center and Observatory, Çanakkale Onsekiz Mart University, Terzioğlu Kampüsü, TR-17020, Çanakkale, Türkiye}

\author{Ahmet KESKİN}
\affil{Astronomy and Space Science, Science Faculty, University of Erciyes, Kayseri 38039, Türkiye}

\author{Mark G. Blackford}
\affil{Variable Stars South (VSS), Congarinni Observatory, Congarinni, NSW, 2447, Australia}

\date{Received ---. revised ---; accepted ---; published}

%\linenumbers
\begin{abstract}
The first multiband photometric solutions of the short-period V Gru eclipsing binary from the southern hemisphere is presented in this study. Light curves of the system were observed through $BVI$ filters at the Congarinni Observatory in Australia for 15 nights. In addition to the new ground-based data, we also used the TESS observations in two sectors. We analyzed the light curves of the system using the PHysics Of Eclipsing BinariEs (PHOEBE) 2.4.7 version code to achieve the best accordance with the photometric observations. The solutions suggest that V Gru is a near-contact binary system with $q=1.302(81)$ mass ratio, $f_1=0.010(23)$, $f_2=-0.0.009(21)$, and $i=73.45(38)$. We considered the two hot spots on the hotter and cooler components for the light curve analysis. We extracted the minima times from the light curves based on the Markov Chain Monte Carlo (MCMC) approach. Using our new light curves, TESS, and additional literature minima, we computed the ephemeris of V Gru. The system's eclipse timing variation trend was determined using the MCMC method. This system is a good and challenging case for future studies.
\end{abstract}


\section{Introduction}
Regarding the type of V Gru system in the binary stars’ categories, contradictory cases are mentioned in the catalogs. In the Washington visual double star catalog (WDS) and the AAVSO international variable star index (VSX), V Gru was classified as a W UMa contact system type (\citealt{2001AJ....122.3466M}). This binary system is classified as a $\beta$ Lyrae in the machine-learned ASAS classification catalog (\citealt{2012ApJS..203...32R}). The review history of this system fully shows its complexity.

\cite{1913HarCi.179....1L} identified V Gru as a new southern variable star with the name of HV 3365. They stated that the HV3365 is an Algol or $\beta$ Lyr type binary system with an apparent magnitude of $V=9.55$. V Gru is indicated by \cite{1978MU...V2..5340H} in the Michigan Catalog of HD stars Vol-2 as HD 207697. \cite{1982IBVS.2185....1W} presented the first minimum time of V Gru. \cite{1983ApJS...52..429W} presented the $V$, $b-y$, $m_1$, $c_1$, and $H_\beta$ values of the V Gru binary system in a photometric survey of the southern hemisphere eclipsing binary stars on nine different Julian days, and they also determined the period of the system as 0.4833 days. \cite{1983A&AS...54..211G} classified this system as a contact or a typical semi-detached configuration, evolved, fairly close binary system. \cite{1984BICDS..27...91B} gave the possible mass ratios in the study, but he stated that the solution is ambiguous. \cite{1993yCat.3135....0C} gave the photovisual and photographic magnitudes of V Gru as 9.80 and 9.50 magnitudes in the Henry Draper Catalogue and Extension. \cite{1996IBVS.4321....1D} gave the new coordinates of V Gru at that time in their study "Accurate Positions of Variable Stars Near the South Galactic Pole". \cite{1998A&AS..129..431H} included V Gru in the $uvby\beta$ photoelectric photometric catalogue. \cite{2000IBVS.4824....1L} stated variables in the HIP, TYC, PPM, and AC catalogues. \cite{2004A&A...417..263B} included V Gru in their catalogue of Algol-type binary stars. \cite{2013AN....334..860A} gave the orbital period of the V Gru binary system as 0.4834454954 days in the Catalogue of Eclipsing Variables (Version 2) and also gave the spectral type and the light curve type as F2V and EW. \cite{2013AJ....146..134K} gave heliocentric radial velocity as $-19.0\pm11.4$ km/s in the RAVE 4th data release (DR4). \cite{2015A-A...580A..23P} gave $V$ magnitude and colour indexes of V Gru in the Stroemgren-Crawford $uvby\beta$ photometry catalog. \cite{2017AJ....153...75K} updated Heliocentric radial velocity as $-18.973\pm11.447$ km/s in the RAVE 5th data release (DR5). The \cite{2019MNRAS.490.3158C} study is included in the Mid-infrared stellar Diameters and Fluxes compilation catalogue (MDF)-version 10. \cite{2020AJ....160...83S} updated the value of heliocentric radial velocity as $-18.95\pm11.34$ km/s in the sixth data release of the Radial Velocity Experiment (RAVE DR6).
\\
\\
In this investigation, we analyze the light curves of the V Gru system. We obtained CCD observations with multi-bands at an observatory in Australia and optimally created binary star models suitable for these observations using the PHOEBE code. The present study is the first in-depth and multiband photometric analysis of this system according to ground-based and space-based observations. Sections present observations, image reduction and instruments used, light curve analysis, the result of photometry, refind ephemeris, light curve behaviour, spectral classification, and finally, conclusion, respectively.

\vspace{1cm}
\section{Observation and data reduction}
\begin{figure*}
\begin{center}
\includegraphics[scale=0.70]{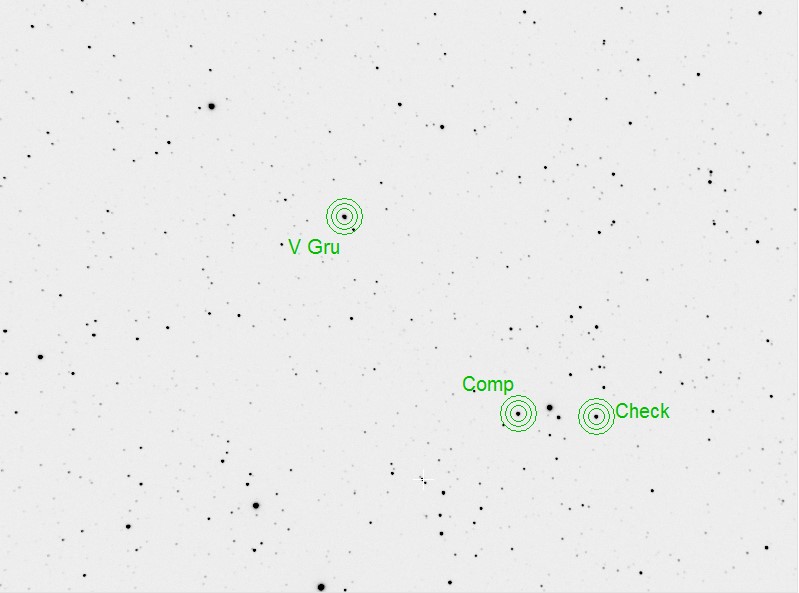}
    \caption{Observed field-of-view for V Gru, comparison star, and check star.}
\label{Fig1}
\end{center}
\end{figure*}

V Gru was observed for 15 nights in August 2020 with an Orion ED80T CF 80mm refractor telescope at the Congarinni Observatory in Australia ({$152^\circ$} $52'$ East and {$30^\circ$} $44'$ South). The data was taken using an Atik One 6.0 CCD camera with $1\times1$ binning and a CCD temperature of $-10^\circ$C. These observations were made using the Astrodon Johnson-Cousins $BVI$ standard filters. TYC 7990-634-1 was selected as a comparison star and TYC 7989-722-1 was chosen as a check star with an appropriate apparent magnitude in comparison to V Gru. The comparison star was found at R.A. $21^h$ $50^m$ $20.93^s$, Dec. {$42^\circ$ $41'$ $46.65"$} (J2000) with a $V=10.291(22)$ magnitude, while the check star was located at R.A. $21^h$ $49^m$ $39.28^s$, Dec. {$-42^\circ$ $42'$ $02.91"$} (J2000) with a $V=10.575(21)$ magnitude, according to the Simbad\footnote{$http://simbad.u-strasbg.fr/simbad$} astronomical database. Figure \ref{Fig1} shows the observed field-of-view for V Gru with the comparison and check stars. During the observations, a total of 1672 images were acquired. The CCD image processing was done with MaxIm DL software, which included dark, bias, and flat-field corrections (\citealt{2000IAPPP..79....2G}).
\\
\\
The Transiting Exoplanet Survey Satellite (TESS) mission observed the V Gru system in addition to our ground-based observations. TESS data is available at the Mikulski Space Telescope Archive (MAST)\footnote{$http://archive.stsci.edu/tess/all\textunderscore products.html$}. The LightKurve code\footnote{$https://docs.lightkurve.org$}  was used to extract TESS style curves from the MAST, which had been detrended using the TESS science processing operations center (SPOC) pipeline (\citealt{2016SPIE.9913E..3EJ}). TESS obtained this data in sectors 1 and 28, which were observed by camera 1 for a total of 55 days.
We used the AstroImageJ (AIJ) program to normalize all of the ground-based and TESS data (\citealt{2017AJ....153...77C}).

\vspace{1cm}
\section{Light curve solutions}
Photometric light curve analysis of the V Gru system was performed by Python Version 2.4.7 of PHOEBE  (\citealt{2005ApJ...628..426P}, \citealt{2020ApJS..250...34C}, \citealt{2022PASP..134f4201P}). We first found a good fit based on all of the initial inputs and the appearance of the light curves. Aside from some initial inputs, the program searched for suitable values for mass ratio, orbital inclination, and star temperature ratio. Then the parameters improved with the optimizing section of the code. We evaluated all binary types to do the light curve analysis and found that only the "semi-detached binary with primary star fills Roche lobe" model produces the best and most logical results.
$g_h=g_c=0.32$ (\citealt{1967ZA.....65...89L}) and $A_h=A_c=0.5$ (\citealt{1969AcA....19..245R}) were used as the bolometric albedo and gravity-darkening coefficients. The stellar atmosphere was described using the \cite{2004A&A...419..725C} approach, and the limb-darkening coefficients were employed as a free parameter in the solutions.

Based on our light curves in the $B$ and $V$ bands and after the necessary calibrations (\citealt{2000A&A...357..367H}), we determined the ($B-V$) of the system to be $0.435\pm0.005$. As a result, the system's effective temperature ($T_1$) was considered to be $6564\pm24$ K (\citealt{1996ApJ...469..355F}). The catalogs list different values of ($B-V$), such as 0.410 (APASS9) and 0.412 (The Hipparcos and Tycho catalogs), with a temperature difference of about 100 K between what was found in this study and what was indicated in the catalogs. We did the initial light curve analysis after fixing the temperature derived from our ($B-V$) for the hotter star. Then, using the optimizing section of PHOEBE, we obtained the temperatures of the stars used in the main photometric light curve solutions.

Using relation 3, the mean fractional radii of the components were computed. Also, we calculated fillout factors from the output parameters of the PHOEBE solutions by relation 4.

\begin{equation}
\label{eq1}\begin{aligned}
r_{mean}=(r_{back}\times r_{side}\times r_{pole})^{1/3}
\end{aligned}
\end{equation}

\begin{equation}
\label{eq2}\begin{aligned}
f=\frac{\Omega(L_1)-\Omega}{\Omega(L_1)-\Omega(L_2)}
\end{aligned}
\end{equation}

\begin{figure*}
\begin{center}
\includegraphics[scale=0.50]{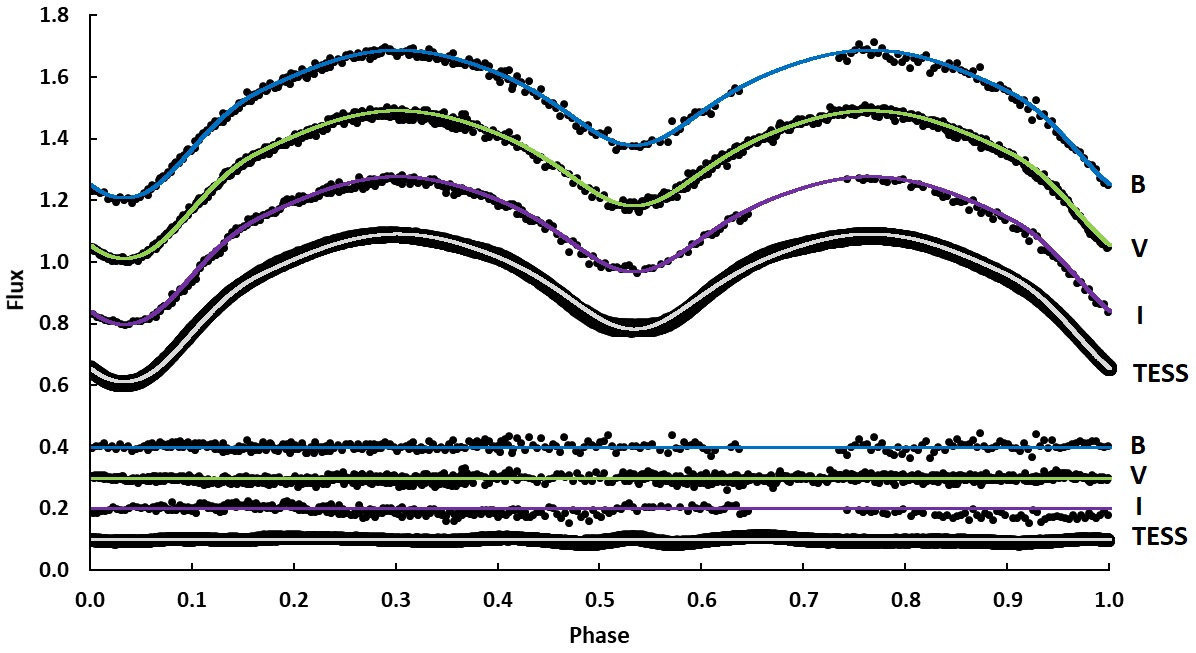}
    \caption{V Gru observational light curves (points) and modeled solutions (lines) are presented with respect to the orbital phase and shifted arbitrarily in the relative flux.}
\label{Fig2}
\end{center}
\end{figure*}

\begin{figure*}
\begin{center}
\includegraphics[scale=0.70]{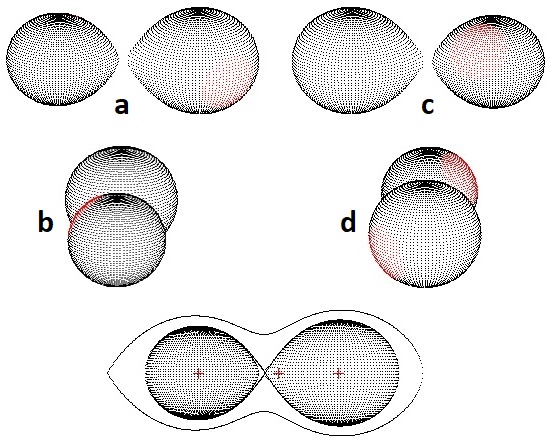}
    \caption{The positions of the components of the V Gru system and the cross-sectional outline of the binary system.}
\label{Fig3}
\end{center}
\end{figure*}

The results of the light curve solutions of the V Gru system are presented in Table \ref{tab1}. The observed and synthetic light curves with residuals are displayed in Figure \ref{Fig2}.
\\
\\
The well-known O'Connell effect (\citealt{1951PRCO....2...85O}), is represented by the asymmetry in the brightness of maxima in the light curve of eclipsing binary systems. The O'Connell effect can be recognized in the light curves from our ground-based observations and TESS data. The light curve solutions in this study need two hot star spots on each of the hooter and cooler components. Figure \ref{Fig3} shows a three-dimensional view of V Gru as well as the Roche lobe configuration.

\begin{table}
\caption{Light curve solutions of V Gru.}
\centering
\begin{center}
\footnotesize
\begin{tabular}{c cc c}
 \hline
 \hline
Parameter &&& Result\\
\hline
$T_{hotter}$ (K) &&& $6749(89)$\\
$T_{cooler}$ (K) &&& $5510(103)$\\
$q=M_2/M_1$ &&& $1.302(81)$\\
$\Omega_{hotter}$ &&& $4.221(66)$\\
$\Omega_{cooler}$ &&& $4.268(47)$\\
$i^{\circ}$ &&&	$73.45(38)$\\
$f_{hotter}$ &&& $0.010(23)$\\
$f_{cooler}$ &&& $-0.009(21)$\\
$A_{hotter}=A_{cooler}$ &&& $0.50$\\
$g_{hotter}=g_{cooler}$ &&& $0.32$\\
$l_{hotter}/l_{tot}$ &&& $0.667(2)$\\
$l_{cooler}/l_{tot}$ &&& $0.333(2)$\\
$r_{hotter}(back)$ &&& $0.430(2)$\\
$r_{hotter}(side)$ &&& $0.399(2)$\\
$r_{hotter}(pole)$ &&& $0.379(2)$\\
$r_{cooler}(back)$ &&& $0.374(3)$\\
$r_{cooler}(side)$ &&& $0.345(3)$\\
$r_{cooler}(pole)$ &&& $0.330(3)$\\
$r_{hotter}(mean)$ &&& $0.402(2)$\\
$r_{cooler}(mean)$ &&& $0.359(2)$\\
Phase shift &&& $0.035(5)$\\
\hline
Hotter star &&&\\
$Colatitude_{spot}(deg)$ &&& 104(2)\\
$Longitude_{spot}(deg)$ &&& 127(2)\\
$Radius_{spot}(deg)$ &&& 32(1)\\
$T_{spot}/T_{star}$ &&& 1.13(1)\\
\hline
Cooler star &&&\\
$Colatitude_{spot}(deg)$ &&& 59(3)\\
$Longitude_{spot}(deg)$ &&& 69(2)\\
$Radius_{spot}(deg)$ &&& 38(1)\\
$T_{spot}/T_{star}$ &&& 1.17(1)\\
\hline
\hline
\end{tabular}
\end{center}
\label{tab1}
\end{table}

\vspace{1cm}
\section{Orbital period variations}
To determine the new times of minima and uncertainties, we utilized Python code using a Gaussian function and the MCMC approach (\citealt{2021NewA...8601571P}). The minimum times in the previous studies and in this study were used to get the potential orbital period change using mid-eclipse times. For this, TIming DAtabase at Krakow (TIDAK) was also taken into account (Table \ref{tab3} and Table A\ref{A1}). In collecting mid-eclipse times in previous reports, we used only those observed with the CCD cameras. All times of minima are expressed in Barycentric Julian Date in Barycentric Dynamical Time ($BJD_{TDB}$).

In this study, a total of seven minima, four primary (Min.$I$) and three secondary (Min.$II$) were obtained separately for V Gru from $BVI$ ground-based observations (Table \ref{tab3}). Since we recorded no color dependence on the timings, we averaged the data acquired from all filters. The reference ephemeris (\citealt{2021JAVSO..49..251R}) is used for calculating epoch and O–C of mid-eclipse times:

\begin{equation}
\label{eq3}\begin{aligned}
BJD_{TDB}(Min.I)=2458727.99696(10)\\
+0.4834461(5)\times E
\end{aligned}
\end{equation}

\begin{figure*}
\begin{center}
\includegraphics[scale=0.40]{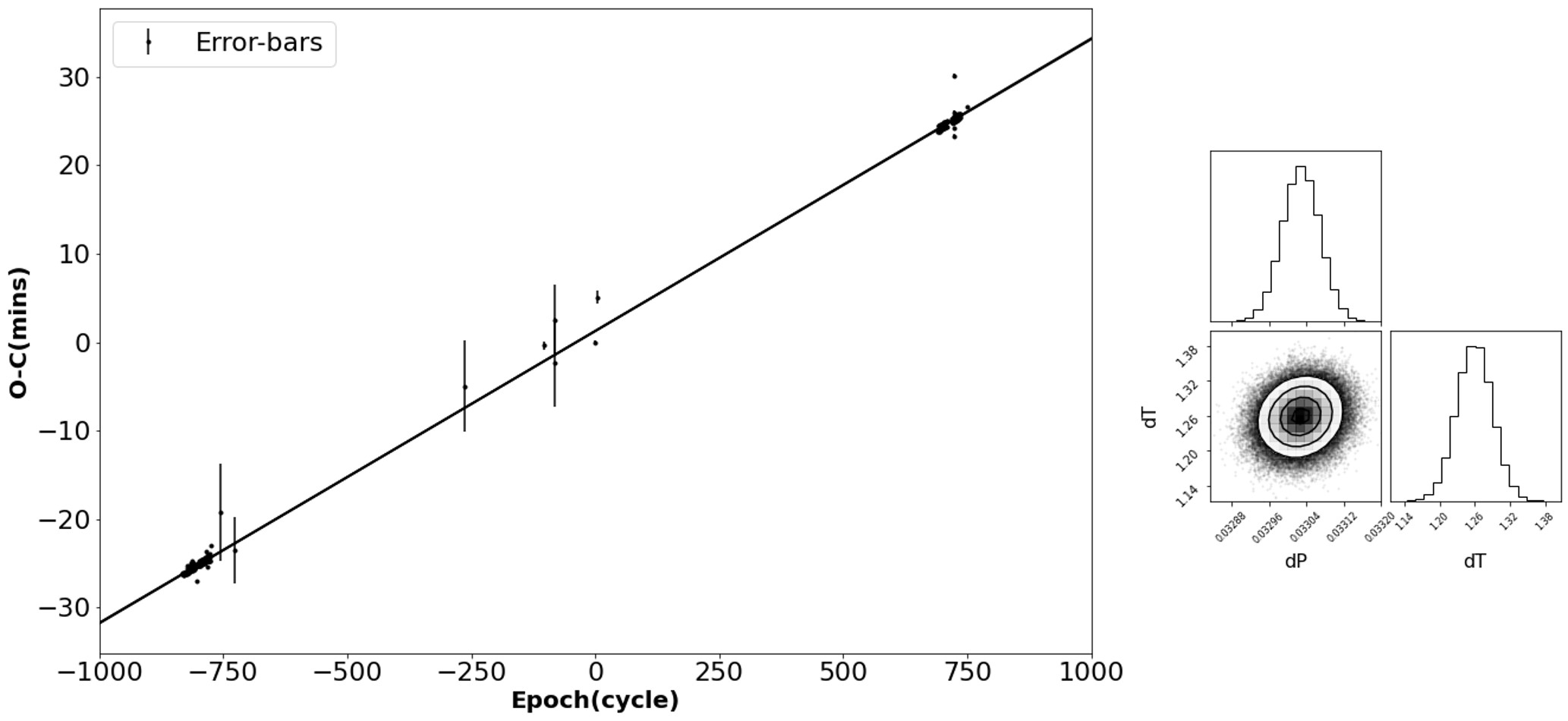}
    \caption{Left: The O-C diagram of V Gru system with the linear fit on the data points. Right: Corner plots of the posterior distribution based on the MCMC sampling.}
\label{Fig4}
\end{center}
\end{figure*}

This study provided a total of 196 mid-eclipse timings for V Gru, including 184 extracted from TESS data, listed in Tables \ref{tab3} and the appendix table. We used the MCMC approach to obtain a new ephemeris for the system (e,g. \citealt{2022PASP..134f4201P}). We find the revised period to be 0.483469038 days for the V Gru system. Figure \ref{Fig4} shows the O-C diagram of the system. A new revised linear ephemeris for the mid-eclipse timings obtained from this study, TESS, and collected from the literature was appointed with the following light elements:

\begin{equation}
\label{eq4}\begin{aligned}
BJD_{TDB}(Min.I)=2458727.99791_{\rm-(4)}^{+(5)}\\
+0.483469038_{\rm-(50)}^{+(52)}\times E 
\end{aligned}
\end{equation}

\begin{table}
\caption{Mid-eclipse times based on $BVI$ observations of this study.}
\centering
\begin{center}
\footnotesize
\begin{tabular}{c c c c c}
 \hline
 \hline
 Min.($BJD_{TDB}$)	& Error	& Filter & Epoch & O-C\\
\hline
2459078.02872	& 0.00016	& $V$	& 724	 & 0.0168\\
2459078.02965	& 0.00016	& $B$	& 724	 & 0.0177\\
2459078.02997	& 0.00017	& $I$	& 724	 & 0.0180\\
2459090.11660	& 0.00004	& $V$	& 749	 & 0.0185\\
2459078.26981	& 0.00021	& $I$	& 724.5 & 0.0162\\
2459078.27130	& 0.00030	& $V$	& 724.5 & 0.0176\\
2459078.27458	& 0.00021	& $B$	& 724.5 & 0.0209\\
\hline
\hline
\end{tabular}
\end{center}
\label{tab3}
\end{table}

\vspace{1cm}
\section{Conclusion}
The V Gru short-period binary system was photometrically observed during a period of 15 nights at a southern hemisphere observatory using $BVI$ filters. We extracted minima times from our observations and TESS data, together with additional minima times from the literature; a new ephemeris for the system was presented using the MCMC approach. The O-C diagram shows an increasing and linear trend.
\\
\\
To analyze the light curves of this system, PHOEBE codes were employed.
The two components have a 1239 K temperature differential, according to Allen's table (\citealt{2000asqu.book.....C}), these temperatures suggest that the spectral types of the hotter and cooler components are F2 and G8, respectively.
\\
\\
The appearance of the light curve is typically the basis for classifications related to machine learning, such as the machine-learned ASAS classification catalog (\citealt{2012ApJS..203...32R}). Classifications are generally designed with the aim of examining systems for which detailed analysis has not yet been provided and appropriate information is not yet available. As a result, they cannot be trusted in systems like V Gru. This may be one of the classifier's faults, which is a problem that artificial intelligence (AI) is hoping to solve in the future. It appears that a different classification offered by \cite{1955AnAp...18..379K} based on the relationship of the components to their surrounding Roche lobes is currently preferred.
\\
\\
The short orbital period and light curve analysis of V Gru shows that this binary system is a near-contact eclipsing binary with the negative fillout factor for the cooler component. So the hotter component has filled its Roche lob, but the cooler star has not yet filled it. However, this system needs investigation with further observations, including spectroscopic analysis.

\vspace{1cm}
\section*{Acknowledgments}
This manuscript was prepared based on a cooperation between the Binary Systems of the South and North Project, the astronomy department of the Raderon AI Lab., and Erciyes University Scientific Research Projects Coordination Unit (Project No. 11737).
In this work, we used $Gaia$ DR3 data from the European Space Agency (ESA) mission $Gaia$ (\url{http://www. cosmos.esa.int/gaia}). The SIMBAD database, operated by CDS, Strasbourg, France, has been used in this study. The National Science Foundation (NSF 1 517 474, 1 909 109) and the National Aeronautics and Space Administration (NASA 17-ADAP17-68) both contributed funding to the PHOEBE that we utilized. We are especially grateful to Prof. Edwin Budding for his valuable comments.

\vspace{1cm}
\section*{ORCID iDs}
\noindent Mehmet TANRIVER: 0000-0002-3263-9680\\
Atila Poro: 0000-0002-0196-9732\\
Ahmet BULUT: 0000-0002-7215-926X\\
Ahmet KESKIN: 0000-0002-9314-0648\\
Mark G. Blackford: 0000-0003-0524-2204\\

\clearpage
\appendix
\setcounter{table}{0}
\begin{table}[h]
\caption{Available mid-eclipse times of V Gru obtained by CCD.}
\centering
\begin{center}
\footnotesize
\begin{tabular}{c c c c c c c c c c}
 \hline
 \hline
Min.($BJD_{TDB}$)	& Error	 & Epoch & O-C & Ref. & Min.($BJD_{TDB}$)	& Error	 & Epoch & O-C & Ref.\\
\hline
2458325.50990	&	0.00002	&	-832.5	&	-0.0182	&	TESS	&	2458335.17979	&	0.00003	&	-812.5	&	-0.0172	&	TESS	\\
2458325.75160	&	0.00001	&	-832	&	-0.0182	&	TESS	&	2458335.42087	&	0.00001	&	-812	&	-0.0179	&	TESS	\\
2458325.99345	&	0.00003	&	-831.5	&	-0.0181	&	TESS	&	2458335.66310	&	0.00003	&	-811.5	&	-0.0173	&	TESS	\\
2458326.23501	&	0.00001	&	-831	&	-0.0182	&	TESS	&	2458335.90437	&	0.00001	&	-811	&	-0.0178	&	TESS	\\
2458326.47681	&	0.00003	&	-830.5	&	-0.0182	&	TESS	&	2458336.14657	&	0.00002	&	-810.5	&	-0.0173	&	TESS	\\
2458326.71852	&	0.00001	&	-830	&	-0.0182	&	TESS	&	2458336.38772	&	0.00001	&	-810	&	-0.0179	&	TESS	\\
2458326.96029	&	0.00002	&	-829.5	&	-0.0181	&	TESS	&	2458336.62999	&	0.00003	&	-809.5	&	-0.0174	&	TESS	\\
2458327.20183	&	0.00002	&	-829	&	-0.0183	&	TESS	&	2458336.87126	&	0.00001	&	-809	&	-0.0178	&	TESS	\\
2458327.44362	&	0.00003	&	-828.5	&	-0.0182	&	TESS	&	2458337.11338	&	0.00003	&	-808.5	&	-0.0174	&	TESS	\\
2458327.68548	&	0.00001	&	-828	&	-0.0181	&	TESS	&	2458337.35473	&	0.00001	&	-808	&	-0.0178	&	TESS	\\
2458327.92716	&	0.00003	&	-827.5	&	-0.0182	&	TESS	&	2458337.59681	&	0.00002	&	-807.5	&	-0.0174	&	TESS	\\
2458328.16892	&	0.00001	&	-827	&	-0.0181	&	TESS	&	2458337.83819	&	0.00002	&	-807	&	-0.0178	&	TESS	\\
2458328.41065	&	0.00002	&	-826.5	&	-0.0181	&	TESS	&	2458338.08006	&	0.00002	&	-806.5	&	-0.0176	&	TESS	\\
2458328.65233	&	0.00001	&	-826	&	-0.0182	&	TESS	&	2458338.32176	&	0.00001	&	-806	&	-0.0176	&	TESS	\\
2458328.89421	&	0.00003	&	-825.5	&	-0.0180	&	TESS	&	2458339.77100	&	0.00001	&	-803	&	-0.0187	&	TESS	\\
2458329.13572	&	0.00001	&	-825	&	-0.0182	&	TESS	&	2458340.01388	&	0.00002	&	-802.5	&	-0.0176	&	TESS	\\
2458329.37761	&	0.00002	&	-824.5	&	-0.0180	&	TESS	&	2458340.25563	&	0.00001	&	-802	&	-0.0176	&	TESS	\\
2458329.61914	&	0.00001	&	-824	&	-0.0182	&	TESS	&	2458340.49740	&	0.00002	&	-801.5	&	-0.0175	&	TESS	\\
2458329.86095	&	0.00003	&	-823.5	&	-0.0181	&	TESS	&	2458340.73909	&	0.00001	&	-801	&	-0.0175	&	TESS	\\
2458330.10266	&	0.00002	&	-823	&	-0.0182	&	TESS	&	2458340.98100	&	0.00003	&	-800.5	&	-0.0174	&	TESS	\\
2458330.34502	&	0.00003	&	-822.5	&	-0.0175	&	TESS	&	2458341.22263	&	0.00001	&	-800	&	-0.0174	&	TESS	\\
2458330.58617	&	0.00001	&	-822	&	-0.0181	&	TESS	&	2458341.46439	&	0.00003	&	-799.5	&	-0.0174	&	TESS	\\
2458330.82825	&	0.00003	&	-821.5	&	-0.0177	&	TESS	&	2458341.70599	&	0.00002	&	-799	&	-0.0175	&	TESS	\\
2458331.06959	&	0.00001	&	-821	&	-0.0181	&	TESS	&	2458341.94791	&	0.00002	&	-798.5	&	-0.0173	&	TESS	\\
2458331.31153	&	0.00002	&	-820.5	&	-0.0179	&	TESS	&	2458342.18946	&	0.00001	&	-798	&	-0.0175	&	TESS	\\
2458331.55305	&	0.00001	&	-820	&	-0.0181	&	TESS	&	2458342.43148	&	0.00003	&	-797.5	&	-0.0172	&	TESS	\\
2458331.79493	&	0.00002	&	-819.5	&	-0.0180	&	TESS	&	2458342.67290	&	0.00001	&	-797	&	-0.0175	&	TESS	\\
2458332.03660	&	0.00002	&	-819	&	-0.0180	&	TESS	&	2458342.91491	&	0.00002	&	-796.5	&	-0.0172	&	TESS	\\
2458332.27842	&	0.00002	&	-818.5	&	-0.0179	&	TESS	&	2458343.15636	&	0.00001	&	-796	&	-0.0175	&	TESS	\\
2458332.52004	&	0.00001	&	-818	&	-0.0180	&	TESS	&	2458343.39840	&	0.00003	&	-795.5	&	-0.0172	&	TESS	\\
2458332.76213	&	0.00003	&	-817.5	&	-0.0176	&	TESS	&	2458343.63983	&	0.00001	&	-795	&	-0.0175	&	TESS	\\
2458333.00348	&	0.00001	&	-817	&	-0.0180	&	TESS	&	2458343.88194	&	0.00003	&	-794.5	&	-0.0171	&	TESS	\\
2458333.24561	&	0.00002	&	-816.5	&	-0.0176	&	TESS	&	2458344.12327	&	0.00001	&	-794	&	-0.0175	&	TESS	\\
2458333.48702	&	0.00001	&	-816	&	-0.0179	&	TESS	&	2458344.36540	&	0.00003	&	-793.5	&	-0.0171	&	TESS	\\
2458333.72912	&	0.00003	&	-815.5	&	-0.0175	&	TESS	&	2458344.60673	&	0.00001	&	-793	&	-0.0175	&	TESS	\\
2458333.97043	&	0.00001	&	-815	&	-0.0180	&	TESS	&	2458344.84880	&	0.00003	&	-792.5	&	-0.0171	&	TESS	\\
2458334.21272	&	0.00003	&	-814.5	&	-0.0174	&	TESS	&	2458345.09025	&	0.00001	&	-792	&	-0.0174	&	TESS	\\
2458334.45392	&	0.00001	&	-814	&	-0.0179	&	TESS	&	2458345.33227	&	0.00003	&	-791.5	&	-0.0171	&	TESS	\\
2458334.69611	&	0.00002	&	-813.5	&	-0.0174	&	TESS	&	2458345.57372	&	0.00001	&	-791	&	-0.0174	&	TESS	\\
2458334.93736	&	0.00001	&	-813	&	-0.0179	&	TESS	&	2458345.81567	&	0.00002	&	-790.5	&	-0.0171	&	TESS	\\
\hline
\hline
\end{tabular}
\end{center}
\label{A1}
\end{table}

\begin{table}[h]
\renewcommand\thetable{1}
\caption{Continued}
\centering
\begin{center}
\footnotesize
\begin{tabular}{c c c c c c c c c c}
 \hline
 \hline
Min.($BJD_{TDB}$)	& Error	 & Epoch & O-C & Ref. & Min.($BJD_{TDB}$)	& Error	 & Epoch & O-C & Ref.\\
\hline
2458346.05722	&	0.00001	&	-790	&	-0.0173	&	TESS	&	2459063.52512	&	0.00001	&	694	&	0.0166	&	TESS	\\
2458346.29922	&	0.00003	&	-789.5	&	-0.0170	&	TESS	&	2459063.76726	&	0.00002	&	694.5	&	0.0170	&	TESS	\\
2458346.54072	&	0.00001	&	-789	&	-0.0173	&	TESS	&	2459064.00865	&	0.00001	&	695	&	0.0167	&	TESS	\\
2458346.78271	&	0.00003	&	-788.5	&	-0.0170	&	TESS	&	2459064.25078	&	0.00002	&	695.5	&	0.0171	&	TESS	\\
2458347.02407	&	0.00001	&	-788	&	-0.0174	&	TESS	&	2459064.49211	&	0.00001	&	696	&	0.0167	&	TESS	\\
2458347.26602	&	0.00003	&	-787.5	&	-0.0171	&	TESS	&	2459064.73422	&	0.00002	&	696.5	&	0.0171	&	TESS	\\
2458347.50800	&	0.00001	&	-787	&	-0.0169	&	TESS	&	2459064.97558	&	0.00001	&	697	&	0.0167	&	TESS	\\
2458347.99100	&	0.00001	&	-786	&	-0.0173	&	TESS	&	2459065.21769	&	0.00002	&	697.5	&	0.0171	&	TESS	\\
2458348.47536	&	0.00005	&	-785	&	-0.0164	&	TESS	&	2459065.45906	&	0.00001	&	698	&	0.0167	&	TESS	\\
2458348.95822	&	0.00005	&	-784	&	-0.0170	&	TESS	&	2459065.70110	&	0.00002	&	698.5	&	0.0170	&	TESS	\\
2458349.44100	&	0.00001	&	-783	&	-0.0177	&	TESS	&	2459065.94251	&	0.00001	&	699	&	0.0167	&	TESS	\\
2458349.68364	&	0.00003	&	-782.5	&	-0.0167	&	TESS	&	2459066.18458	&	0.00002	&	699.5	&	0.0171	&	TESS	\\
2458349.92493	&	0.00001	&	-782	&	-0.0172	&	TESS	&	2459066.42605	&	0.00001	&	700	&	0.0168	&	TESS	\\
2458350.16688	&	0.00003	&	-781.5	&	-0.0170	&	TESS	&	2459066.66808	&	0.00002	&	700.5	&	0.0171	&	TESS	\\
2458350.40842	&	0.00001	&	-781	&	-0.0171	&	TESS	&	2459066.90959	&	0.00001	&	701	&	0.0169	&	TESS	\\
2458350.65038	&	0.00003	&	-780.5	&	-0.0169	&	TESS	&	2459067.15163	&	0.00002	&	701.5	&	0.0172	&	TESS	\\
2458350.89183	&	0.00001	&	-780	&	-0.0172	&	TESS	&	2459067.39294	&	0.00001	&	702	&	0.0168	&	TESS	\\
2458351.13379	&	0.00002	&	-779.5	&	-0.0169	&	TESS	&	2459067.63503	&	0.00002	&	702.5	&	0.0172	&	TESS	\\
2458351.37526	&	0.00001	&	-779	&	-0.0172	&	TESS	&	2459067.87636	&	0.00001	&	703	&	0.0168	&	TESS	\\
2458351.61747	&	0.00003	&	-778.5	&	-0.0167	&	TESS	&	2459068.11850	&	0.00002	&	703.5	&	0.0172	&	TESS	\\
2458351.85874	&	0.00001	&	-778	&	-0.0172	&	TESS	&	2459068.35982	&	0.00001	&	704	&	0.0168	&	TESS	\\
2458352.10074	&	0.00003	&	-777.5	&	-0.0169	&	TESS	&	2459068.60201	&	0.00002	&	704.5	&	0.0173	&	TESS	\\
2458352.34218	&	0.00001	&	-777	&	-0.0172	&	TESS	&	2459068.84332	&	0.00001	&	705	&	0.0169	&	TESS	\\
2458352.58441	&	0.00003	&	-776.5	&	-0.0167	&	TESS	&	2459069.08546	&	0.00002	&	705.5	&	0.0173	&	TESS	\\
2458352.82562	&	0.00001	&	-776	&	-0.0172	&	TESS	&	2459069.32683	&	0.00001	&	706	&	0.0169	&	TESS	\\
2458353.06853	&	0.00003	&	-775.5	&	-0.0160	&	TESS	&	2459069.56887	&	0.00002	&	706.5	&	0.0172	&	TESS	\\
2458362.98178	&	0.00380	&	-755	&	-0.0134	&	\cite{2021JAVSO..49..251R}	&	2459069.81025	&	0.00001	&	707	&	0.0169	&	TESS	\\
2458376.99878	&	0.00260	&	-726	&	-0.0163	&	\cite{2021JAVSO..49..251R}	&	2459070.05236	&	0.00002	&	707.5	&	0.0173	&	TESS	\\
2458600.84718	&	0.00360	&	-263	&	-0.0035	&	TIDAK	&	2459070.29378	&	0.00001	&	708	&	0.0170	&	TESS	\\
2458678.20175	&	0.00031	&	-103	&	-0.0003	&	\cite{2021JAVSO..49..251R}	&	2459070.53592	&	0.00002	&	708.5	&	0.0174	&	TESS	\\
2458688.11438	&	0.00280	&	-82.5	&	0.0017	&	TIDAK	&	2459070.77718	&	0.00001	&	709	&	0.0169	&	TESS	\\
2458688.35277	&	0.00350	&	-82	&	-0.0016	&	TIDAK	&	2459071.01930	&	0.00002	&	709.5	&	0.0173	&	TESS	\\
2458727.99696	&	0.00010	&	0	&	0	&	\cite{2021JAVSO..49..251R}	&	2459075.37048	&	0.00002	&	718.5	&	0.0175	&	TESS	\\
2458730.17597	&	0.00052	&	4.5	&	0.0035	&	\cite{2020JAVSO..48..250R}	&	2459075.61194	&	0.00001	&	719	&	0.0172	&	TESS	\\
2459062.07470	&	0.00001	&	691	&	0.0165	&	TESS	&	2459075.85395	&	0.00002	&	719.5	&	0.0175	&	TESS	\\
2459062.31694	&	0.00002	&	691.5	&	0.0170	&	TESS	&	2459076.09541	&	0.00001	&	720	&	0.0173	&	TESS	\\
2459062.55821	&	0.00001	&	692	&	0.0165	&	TESS	&	2459076.33734	&	0.00002	&	720.5	&	0.0175	&	TESS	\\
2459062.80034	&	0.00002	&	692.5	&	0.0170	&	TESS	&	2459076.57882	&	0.00001	&	721	&	0.0172	&	TESS	\\
2459063.04162	&	0.00001	&	693	&	0.0165	&	TESS	&	2459076.82085	&	0.00002	&	721.5	&	0.0175	&	TESS	\\
2459063.28382	&	0.00002	&	693.5	&	0.0170	&	TESS	&	2459077.06236	&	0.00001	&	722	&	0.0173	&	TESS	\\
\hline
\hline
\end{tabular}
\end{center}
\label{A2}
\end{table}

\begin{table}[h]
\renewcommand\thetable{1}
\caption{Continued}
\centering
\begin{center}
\footnotesize
\begin{tabular}{c c c c c c c c c c}
 \hline
 \hline
Min.($BJD_{TDB}$)	& Error	 & Epoch & O-C & Ref. & Min.($BJD_{TDB}$)	& Error	 & Epoch & O-C & Ref.\\
\hline
2459077.30438	&	0.00002	&	722.5	&	0.0176	&	TESS	&	2459080.93009	&	0.00001	&	730	&	0.0175	&	TESS	\\
2459077.54585	&	0.00001	&	723	&	0.0174	&	TESS	&	2459081.17185	&	0.00002	&	730.5	&	0.0175	&	TESS	\\
2459077.78791	&	0.00002	&	723.5	&	0.0177	&	TESS	&	2459081.41355	&	0.00001	&	731	&	0.0175	&	TESS	\\
2459078.02924	&	0.00001	&	724	&	0.0173	&	TESS	&	2459081.65566	&	0.00002	&	731.5	&	0.0179	&	TESS	\\
2459078.27126	&	0.00002	&	724.5	&	0.0176	&	TESS	&	2459081.89704	&	0.00001	&	732	&	0.0175	&	TESS	\\
2459078.51272	&	0.00001	&	725	&	0.0173	&	TESS	&	2459082.13905	&	0.00002	&	732.5	&	0.0178	&	TESS	\\
2459078.75480	&	0.00002	&	725.5	&	0.0177	&	TESS	&	2459082.38055	&	0.00001	&	733	&	0.0176	&	TESS	\\
2459078.99621	&	0.00001	&	726	&	0.0174	&	TESS	&	2459082.62257	&	0.00002	&	733.5	&	0.0179	&	TESS	\\
2459079.23824	&	0.00002	&	726.5	&	0.0177	&	TESS	&	2459082.86399	&	0.00001	&	734	&	0.0176	&	TESS	\\
2459079.47966	&	0.00001	&	727	&	0.0174	&	TESS	&	2459083.10602	&	0.00002	&	734.5	&	0.0179	&	TESS	\\
2459079.72164	&	0.00002	&	727.5	&	0.0176	&	TESS	&	2459083.34745	&	0.00001	&	735	&	0.0176	&	TESS	\\
2459079.96311	&	0.00001	&	728	&	0.0174	&	TESS	&	2459083.58949	&	0.00002	&	735.5	&	0.0179	&	TESS	\\
2459080.20524	&	0.00002	&	728.5	&	0.0178	&	TESS	&	2459083.83091	&	0.00001	&	736	&	0.0176	&	TESS	\\
2459080.44660	&	0.00001	&	729	&	0.0174	&	TESS	&	2459084.07287	&	0.00002	&	736.5	&	0.0179	&	TESS	\\
2459080.68867	&	0.00002	&	729.5	&	0.0178	&	TESS	&		&		&		&		&		\\
\hline
\hline
\end{tabular}
\end{center}
\label{A3}
\end{table}

\clearpage
\bibliographystyle{aasjournal}
\bibliography{new.ms}

\end{document}